\documentclass[12pt]{article}
\usepackage{graphicx}
\usepackage{amsmath}
\usepackage{bm}
\def\Vec#1{\mbox{\boldmath $#1$}}

\def\itmb{\begin{itemize}}
\def\itme{\end{itemize}}
\def\enmb{\begin{enumerate}}
\def\enme{\end{enumerate}}
\def\eqnb{\begin{equation}}
\def\eqne{\end{equation}}

\def\PRL{Phys. Rev. Lett.}
\def\PRD{{Phys. Rev.} D}

\def\RMP{{Rev. Mod. Phys.}}

\title{Cartan's Supersymmetry and \\the violation of $CP$ symmetry}  

\author{Sadataka Furui\\
 Graduate School of Teikyo University\\
2-17-12 Toyosatodai, Utsunomiya, 320-0003 Japan,\\
 e-mail furui@umb.teikyo-u.ac.jp
}
\begin{document}
\maketitle
\begin{abstract}
 Cartan's supersymmetriy fixes couplings of two types of fermions $\psi, \phi$ and two types of vector fields $x_i$ and $x_{i'}$, ($i=1,2,3,4$). 
The electromagnetic interaction of leptons and quarks is expressed as ${^t\psi}{\mathcal C}x_i\psi$.
In the case of coupling of leptons and quarks with $W$, we extend the coupling ${^t\phi}{\mathcal C}X\psi$ to ${^t\phi}{\mathcal C}X(1-\gamma_5)\psi$, where $X=x_i$ or ${x_i}'$, and unify the interactions in the form $
{^t\phi}{\mathcal C}\bar{x_i}\psi+{^t\phi}{\mathcal C} \bar{{x_i}'} {\mathcal C}\psi 
$, 
where $\bar{x_i}$ implies appropriate $x_i$ or $(-\gamma_5 x_i)$ is chosen. The $\gamma_5$
term induces in time component of $B^0\to K^0\, J/\Psi$,  $K^0$ with the $\bar s$ quark dominated by the small component, while in space components of $\bar B^0\to \bar K^0\, J/\Psi$,  $\bar K^0$ with the $s$ quark dominated by the small component. This asymmetry could be the origin of the $CP$ asymmetry in the $B^0\to K^0\, J/\Psi$.

We apply the model to $B_{s(d)}^0\to D_{s(d)}^- D_{s(d)}^+$ decays and study violation of  $CP$ symmetry via interference of tree and penguin contributions, and observe that the asymmetry is weak consistent to the experiment.
\end{abstract}


\section{Introduction}
Violation of $CP$ symmetry was observed in the difference of  $B^0\to K^0+X $ decay and $\bar B^0\to \bar K^0+X$ decay, where data with $X$ chosen to be $J/\Psi$ or $\pi^\pm$ \cite{BeMa15,Babar09}. In the experiment of producing $B^0, \bar B^0$ entangled state in the $\Upsilon(4S)$ resonance decay, conditions of $B^0$ or $\bar B^0$ decay into $\ell^+ X$ or $\ell^- X$ and another  $B^0$ or $\bar B^0$ decay into $J/\Psi K^0$ or $J/\Psi \bar K^0$  were chosen.

We consider violation of $CP$ symmetry or time reversal symmetry in the decay of $B$ mesons observed in the difference of
\[
B^0\to B_+ (\ell^-\nu_{\ell}+ X, J/\Psi\bar K_L^0 ) \, {\rm v.s.}\, \bar B^0\to B_- ( \ell^+\nu_{\ell}+ X, J/\Psi K_L^0 ). 
\]
where $K_S^0$ and $K_L^0$ are linear combinations of $K^0$ and $\bar K^0$,
\begin{eqnarray}
 |K_S^0\rangle={\mathcal N}(p_k |K^0\rangle- q_k |\bar K^0\rangle)\nonumber\\
 |K_L^0\rangle={\mathcal N}(p_k |K^0\rangle+ q_k |\bar K^0\rangle)\nonumber,
\end{eqnarray}
where $\mathcal N=1/\sqrt{p_k^2+q_k^2}$.

From the difference of experimental data of
\[
B^0\to \pi^- K^+  \, {\rm v.s.}\, \bar B^0\to  \pi^+ K^-
\]
one can evaluate the CKM angle $\gamma$ and $CP$ violation\cite{FM97,GNPS97}, but we do not take this approach, since an entangled neutral $B$ meson state $\Upsilon(4S)$ affects the phase $\gamma$\cite{BeMa15}.  

\begin{figure}
\begin{minipage}[b]{0.47\linewidth}
\begin{center} 
\includegraphics[width=6cm,angle=0,clip]{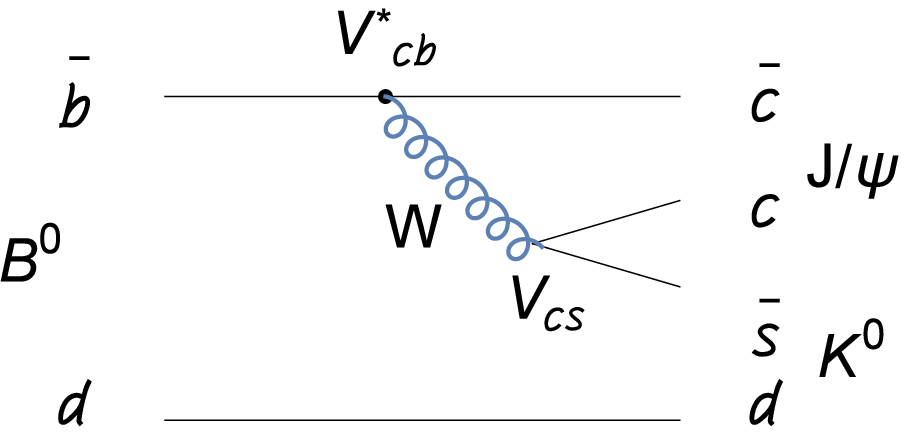} 
\end{center}
\caption{Typical tree diagram of  $B^0\to K_L^0\, J/\Psi$ decay. $V_{cb}$ and $V_{cs}$ are CKM amplitudes.}
\label{B0KT}
\end{minipage}
\hfill
\begin{minipage}[b]{0.47\linewidth}
\begin{center}
\includegraphics[width=6cm,angle=0,clip]{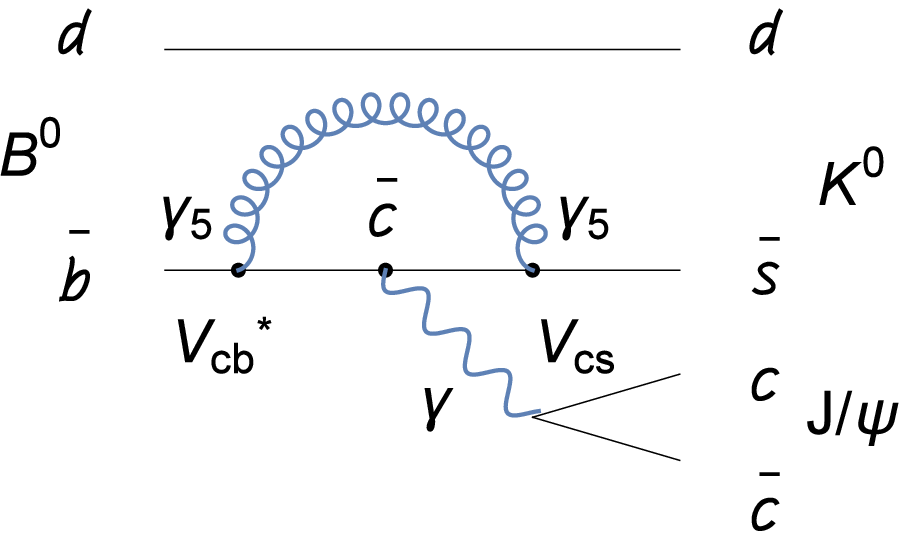}  
\end{center}
\caption{Typical penguin diagram of  $B^0\to K_L^0\, J/\Psi$ decay. In our model we include the loop of $\gamma$s in addition to that of $W$ bosons.}
\label{B0KP}
\end{minipage}
\end{figure}

Cartan's supersymmetry\cite{Cartan66} defines the interaction of  spinors ( quarks or anti-quarks in $B^0$ meson) and vector particles ($W, Z$ bosons or $\gamma$ particles). In our application of the model to the electro-magnetic decay of a Higgs particle ($H^0\to\gamma\gamma$) and weak decay ($H^0\to W\bar W$) \cite{SF12a,SF14,SF15,SF15a} suggest that the quark-gluon and the quark-photon interaction of Cartan's supersymmetry could give clear signal of violation of the $CP$ symmetry.
In the standard model, $CP$ symmetry violation occurs from the interference of the tree diagram amplitude and the penguin diagram amplitude\cite{Babar09}. Typical tree diagrams and penguin diagrams of this decay are shown in Fig.\ref{B0KT} and Fig.\ref{B0KP}.

The decay rates of $\bar B^0(B^0)\to f$ is parametrized as
\begin{equation}
\Gamma_{\bar B^0(B^0)\to f}\propto e^{-\Gamma|\Delta t|}\{ 1+(-)[S_f\sin(\Delta m_d\Delta t) -C_f\cos(\Delta m_d\Delta t)]\},\nonumber
\end{equation}
where $\Delta t=t_{CP}-t_{flavor}$ is the difference of the decay time to the $CP$ eigenstates and the decay time to the flavor eigenstates, and $\Delta m_d$ is the mass difference between the physical states of the neutral $B$ meson systems. The decay time difference $\Delta t$ is approximately given by the distance $\Delta d$ between the decay points and the Lorentz boost $\beta\gamma$ as
$\Delta t\simeq \Delta d/\beta\gamma c$. 
 
Adopting the parametrization of the low mass eigenstate $B_L$ and the high mass eigenstate $B_H$ as
\begin{eqnarray}
|B_L\rangle&\propto p\sqrt{1-z}|B^0\rangle -q\sqrt{1+z}|\bar B^0\rangle\nonumber\\ 
|B_H\rangle&\propto p\sqrt{1+z}|B^0\rangle +q\sqrt{1-z}|\bar B^0\rangle\nonumber,
\end{eqnarray} 
where $|q/p|=1$ and $z=0$, if symmetry holds.
The amplitude $S_f$ and $C_f$ are parametrized as
\[
S_f=2 Im \lambda_f/(1+|\lambda_f|^2),\quad C_f=(1-|\lambda_f|^2)/(1+|\lambda_f|^2),
\]
where using the $CP$ parity $\eta_f=-1(+1)$ for $f =J/\Psi K_S^0(J/\Psi K_L^0)$ and 
\[
\lambda_f=\eta_f (q/p)(\bar A_f/A_f)(p_K/q_k). 
\]
Here $\bar A_f/A_f$ is the ratio of the $\bar B^0\to f$ and $B^0\to f$ decays. 
 
Experimentally, violation of $CP$ or time reversal symmetry can be observed via difference of $\Delta t$ dependence of $B^0$ tagged events and $\bar B^0$ tagged events\cite{BeMa15,Babar09}. The difference of $B^0$ tagged events and $\bar B^0$ tagged events shows the direction of the time axis\cite{Zeller12}.

The propagator between vertices $V^*_{cb}$ and $V_{cs}$ is assigned as  a $W$ boson, and the source of $c\bar c$ in the penguin diagram is assigned as a $\gamma$ photon or $Z$ boson.   
In our model, a $\gamma$ particle $A_\mu$ is emitted from a quark in $B^0$ or $\bar B^0$ as
\[
\bar\psi_L( \gamma_\mu( i\partial_\mu-e\,A_\mu)-m)\psi_L.
\]
The quark $\psi_L$ which consists of $\psi$ and its small component ${\mathcal C}\psi$ interacts with an anti-quark $\phi_L$ which consists of $\phi$ and its small component ${\mathcal C}\phi$. A $\gamma$ particle $A_\mu$ is emitted also from $\phi_L$ as
\[
\bar\phi_L( \gamma_\mu( i\partial_\mu-e\,A_\mu)-m)\phi_L.
\]
 
In the $SU(2)\times U(1)$ theory, $A_\mu$, $Z_\mu$ and $W_\mu$ are related to the gauge fields $A_\mu'$ and $B_\mu'$ by
\begin{eqnarray}
A_\mu=\sin\theta_W {A'_\mu}^3+\cos\theta_W B_\mu'\nonumber\\
Z_\mu=\cos\theta_W {A'_\mu}^3-\sin\theta_W B_\mu'\nonumber\\
W_\mu^\pm=({A'_\mu}^1\mp i{A'_\mu}^2)/\sqrt 2\nonumber,
\end{eqnarray}
where $\theta_W$ is the Weinberg angle.

In the standard model, the weak interaction is described as\cite{BBJ81}
\begin{eqnarray}
&&{\bar\psi_L}^k \gamma_\mu (i\partial_\mu+g_2\frac{\tau^a}{2}{W_0^a}_\mu+g_1\frac{y_k}{2}B_\mu){\psi_L^k}\nonumber\\
&&+{\bar a^k}_R \gamma^\mu (i\partial_\mu+g_1\frac{1+y_k}{2}B_\mu) {a_R^k}\nonumber\\
&&+{\bar b^k}_R \gamma^\mu (i\partial_\mu+g_1\frac{-1+y_k}{2}B_\mu) {b_R^k},
\end{eqnarray}
where $a^k_R$ and $b^k_R$ are right-handed spinors. 
The weak isospin of quarks or leptons are defined by $y^k$. The isospin of $W_0^a$ and $B_\mu$ are rotated, and physical states $W_\mu$, $A_\mu$, $Z_\mu$ and $\psi_L$ appear. 
We expect that physical $\phi_L$ appear together with the physical $\psi_L$.

The Cabibbo-Kobayashi-Maskawa (CKM) amplitudes  $V_{cs}, V_{cb}, V_{cd}$ are parametrized as
\begin{eqnarray}
V_{cs}&=&1-\frac{1}{2}\lambda^2-\frac{1}{8}\lambda^4(1+4A^2)\nonumber\\
V_{cb}&=&A\lambda^2\nonumber\\
V_{cd}&=&-\lambda+\frac{1}{2}A^2\lambda^5[1-2(\rho+i\eta)]\nonumber
\end{eqnarray} 
where $\lambda\simeq 0.23$.

Large $V_{cs}=0.973$ yields strong $B^0(\bar B^0)\to K^0+X (\bar K^0+X)$ decay modes. The decay mode $B^0(\bar B^0)\to D^{*-}\ell^+\nu_\ell( D^{*+}\ell^-\nu_\ell)$ is strong due to relatively large $V_{cd}=0.225$ and small $V_{ub}=0.00355$ and $V_{cb}=0.0414$.
The  relation $V_{cb}=A\lambda^2=0.0414$ requires $A=0.78$, while $V_{cs}=0.973$ requires $A=0.38$, and the parametrization of the CKM amplitudes are not completed.
We study also the decay of $B^0(\bar B^0)\to D^+ D^-$ and $B_s(\bar B_s) \to D_s^+D_s^-$.

In section 2, we explain Cartan's supersymmetry in $B^0\to K^0\,J/\Psi$ interaction, and in section3, we explain differences of $B^0\to D^+D^-$ and $B_s\to D_s^+D_s^-$ from Cartan's supersymmetry.

\section{Cartan's supersymmetry and weak interactions}
We want to reproduce the qualitative features of CKM amplitudes from the trialty model.
In the previous work\cite{SF15}, we could understand absence of $B_s(0^+)^+\to B_s(0^-)\pi^+$, and presence of $B_s(0^+)^+\to D_s^*(0^+)^+\to D_s(0^-)\pi^+$ from the  triality selection rules that Cartan's supersymmetry predicts\cite{Cartan66,SF12a,SF14}. 

We define a Dirac spinor composed of $\psi$ and  ${\mathcal C}\psi$
\begin{eqnarray}
\psi&=&\xi_1\Vec i+\xi_2\Vec j+\xi_3\Vec k+\xi_4\Vec I\nonumber\\
&=&\left(\begin{array}{cc}\xi_4+i\xi_3 &i \xi_1-\xi_2\\
                                i\xi_1+\xi_2&\xi_4-i\xi_3\end{array}\right)\nonumber\\
{\mathcal C}\psi&=&-\xi_{234}\Vec i-\xi_{314}\Vec j-\xi_{124}\Vec k+\xi_{123}\Vec I\nonumber\\
&=&\left(\begin{array}{cc}\xi_{123}-i\xi_{124}&-i\xi_{234}+\xi_{314}\\
                                -i\xi_{234}-\xi_{314}&\xi_{123}+i\xi_{124}\end{array}\right)
\end{eqnarray}
and another Dirac spinor composed of $\phi$ and ${\mathcal C}\phi$
\begin{eqnarray}
\phi&=&\xi_{14}\Vec i+\xi_{24}\Vec j+\xi_{34}\Vec k+\xi_0\Vec I\nonumber\\
&=&\left(\begin{array}{cc}\xi_0+i\xi_{34} &i \xi_{14}-\xi_{24}\\
                                i\xi_{14}+\xi_{24}&\xi_0-i\xi_{34}\end{array}\right)\nonumber\\
{\mathcal C}\phi&=&-\xi_{23}\Vec i-\xi_{31}\Vec j-\xi_{12}\Vec k+\xi_{1234}\Vec I\nonumber\\
&=&\left(\begin{array}{cc}\xi_{1234}-i\xi_{12} &-i \xi_{23}+\xi_{31}\\
                                -i\xi_{23}-\xi_{31}&\xi_{1234}+i\xi_{12}\end{array}\right)
\end{eqnarray}
which interact with four dimensional vector fields $X$.

The trilinear form of electromagnetic interaction in these bases is
\begin{eqnarray}
{\mathcal F}&=&^t\phi {\mathcal C}X\psi={^t\phi} \gamma_0x^\mu\gamma_\mu\psi\nonumber\\
&=&x_1(\xi_{12}\xi_{314}-\xi_{31}\xi_{124}-\xi_{14}\xi_{123}+\xi_{1234}\xi_1)\nonumber\\
&+&x_2(\xi_{23}\xi_{124}-\xi_{12}\xi_{234}-\xi_{24}\xi_{123}+\xi_{1234}\xi_2)\nonumber\\
&+&x_3(\xi_{31}\xi_{234}-\xi_{23}\xi_{314}-\xi_{34}\xi_{123}+\xi_{1234}\xi_3)\nonumber\\
&+&x_4(-\xi_{14}\xi_{234}-\xi_{24}\xi_{314}-\xi_{34}\xi_{124}+\xi_{1234}\xi_4)\nonumber\\
&+&x_{1'}(-\xi_{0}\xi_{234}+\xi_{23}\xi_{4}-\xi_{24}\xi_{3}+\xi_{34}\xi_2)\nonumber\\
&+&x_{2'}(-\xi_{0}\xi_{314}+\xi_{31}\xi_{4}-\xi_{34}\xi_{1}+\xi_{14}\xi_3)\nonumber\\
&+&x_{3'}(-\xi_{0}\xi_{124}+\xi_{12}\xi_{4}-\xi_{14}\xi_{2}+\xi_{24}\xi_1)\nonumber\\
&+&x_{4'}(\xi_{0}\xi_{123}-\xi_{23}\xi_{1}-\xi_{31}\xi_{2}-\xi_{12}\xi_3)
\end{eqnarray}

In the case of weak interaction, we replace the coupling $\gamma_0x^\mu\gamma_\mu$ to
$\gamma_0x^\mu\gamma_\mu(1-\gamma_5)$ and try to make the couplings between fermions and vector particles become unified in the form
\[
\sum_{i=1}^4 (x_i {\mathcal C}\phi\, {\mathcal C}\psi +x_{i'}{\mathcal C}\phi\, \psi)
\]
by suitable choice of $1$ or $-\gamma_5$. 

Except the term $x_{4'}\xi_0\xi_{123}$, which is $x_{i'} \phi {\mathcal C}\psi$ type, it is possible by the following choice
\begin{eqnarray}
{\mathcal G}&=&x_1(\xi_{12}\xi_{314}-\xi_{31}\xi_{124}+\langle\xi_{14}\gamma_5\rangle\xi_{123}-\xi_{1234}\langle\gamma_5\xi_1\rangle)\nonumber\\
&+&x_2(\xi_{23}\xi_{124}-\xi_{12}\xi_{234}+\langle\xi_{24}\gamma_5\rangle\xi_{123}-\xi_{1234}\langle\gamma_5\xi_2\rangle)\nonumber\\
&+&x_3(\xi_{31}\xi_{234}-\xi_{23}\xi_{314}+\langle\xi_{34}\gamma_5\rangle\xi_{123}-\xi_{1234}\langle\gamma_5\xi_3\rangle)\nonumber\\
&+&x_4(\langle\xi_{14}\gamma_5\rangle\xi_{234}+\langle\xi_{24}\gamma_5\rangle\xi_{314}+\langle\xi_{34}\gamma_5\rangle\xi_{124}-\xi_{1234}\langle\gamma_5\xi_4\rangle)\nonumber\\
&+&x_{1'}(\langle\xi_{0}\gamma_5\rangle\xi_{234}+\xi_{23}\xi_{4}-\xi_{24}\xi_{3}+\langle - \xi_{34}\gamma_5\rangle\xi_2)\nonumber\\
&+&x_{2'}(\langle\xi_{0}\gamma_5\rangle\xi_{314}+\xi_{31}\xi_{4}-\xi_{34}\xi_{1}+\langle- \xi_{14}\gamma_5\rangle\xi_3)\nonumber\\
&+&x_{3'}(\langle\xi_0 \gamma_5\rangle\xi_{124}+\xi_{12}\xi_{4}-\xi_{14}\xi_{2}+\langle- \xi_{24}\gamma_5\rangle\xi_1)\nonumber\\
&+&x_{4'}(\xi_0 \xi_{123}-\xi_{23}\xi_{1}-\xi_{31}\xi_{2}-\xi_{12}\xi_3).
\end{eqnarray}
We allow also direct channel couplings of $^t\phi \mathcal C{x_4}'\mathcal C\phi$ and $^t\psi {x_4}'\psi$. 

Experimentally, violation of $CP$ was observed in the decay of $B^0$ and $B_s$ around 1997. In the case of $B\to K\pi$, there is a constraint on the unitarity angle $\gamma$, which is called Fleischer-Mannel bound by Grossmann et al\cite{FM97,GNPS97}. They derived the bound for the angle $\gamma$ of unitary triangle, $\sin^2\gamma>R+1.645\sigma_R$, with $R=0.65+/- 0.40$ and the uncertainty $\sigma_R$. 

In comparison to $B\to K\pi$ decay, $B^0\to K^0 J/\Psi$ contains weak final state interactions.
In order to evaluate the $B^0\to K^0 J/\Psi$ decay amplitudes we fix the initial configuration of $B^0=d\bar b$ and the final configuration $K^0=d\bar s$, $J/\Psi=c\bar c$, and a quark is expressed as a Dirac spinor ${^t(}\psi,\mathcal C \psi)$ and an anti-quark is expressed as a Dirac spinor ${^t(}\phi,\mathcal C \phi)$.

\begin{figure}
\begin{minipage}[b]{0.47\linewidth}
\begin{center}
\includegraphics[width=6cm,angle=0,clip]{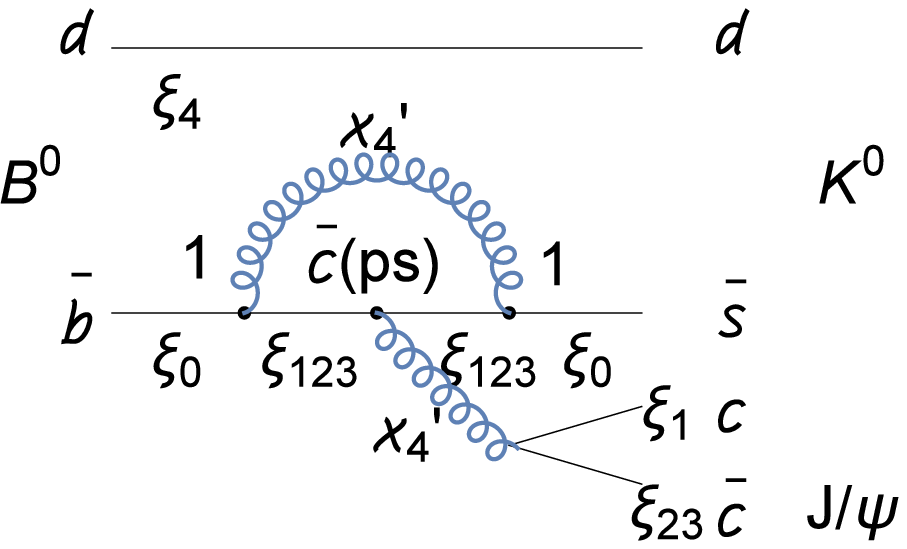} 
\end{center}
\label{bscc55b}
\end{minipage}
\hfill
\begin{minipage}[b]{0.47\linewidth}
\begin{center} 
\includegraphics[width=6cm,angle=0,clip]{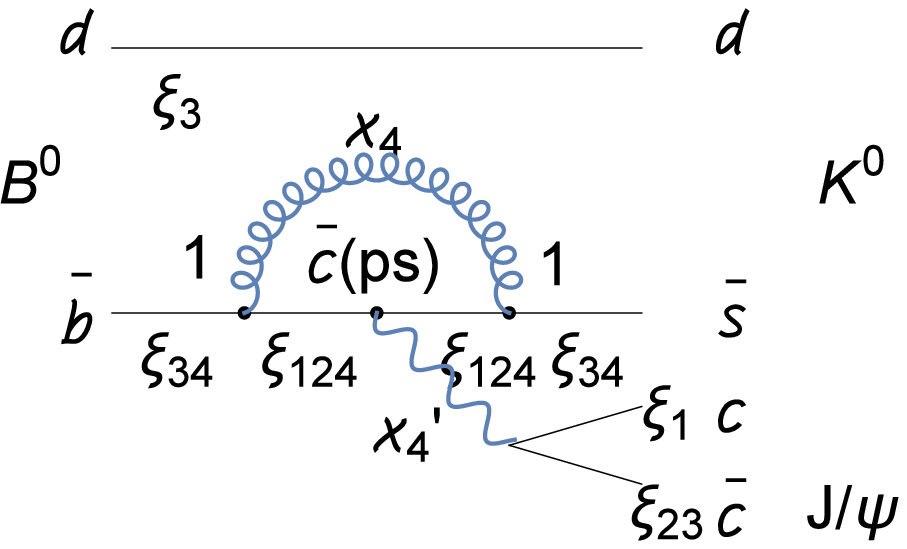}
\end{center}
\label{bscc55a1}
\end{minipage}
\begin{minipage}[b]{0.47\linewidth}
\begin{center}
\includegraphics[width=6cm,angle=0,clip]{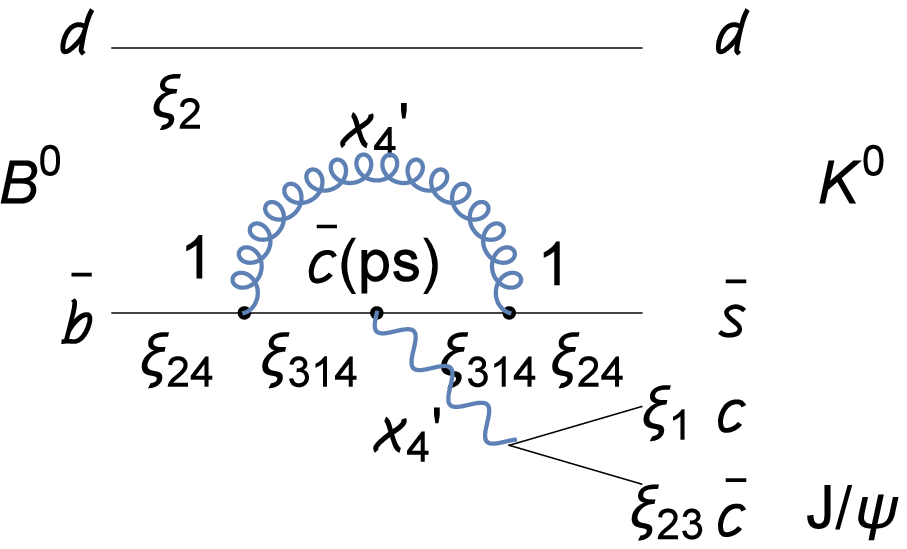} 
\end{center}
\label{bscc55a}
\end{minipage}
\hfill
\begin{minipage}[b]{0.47\linewidth}
\begin{center} 
\includegraphics[width=6cm,angle=0,clip]{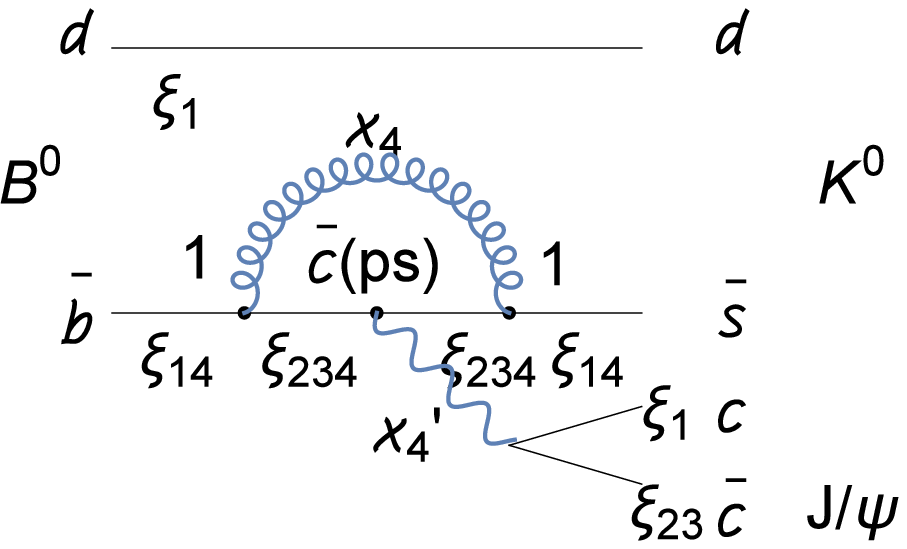} 
\end{center}
\end{minipage}
\caption{Typical diagrams of  $B^0\to K^0\, J/\Psi$ decay, penguin diagram $1\, 1$ type. In the $J/\Psi$ configurations, ${x_4}'$ decays to $\xi_{1}\xi_{23}$, $\xi_{2}\xi_{31}$ or $\xi_{3}\xi_{12}$.}
\label{bscc11}
\begin{minipage}[b]{0.47\linewidth}
\begin{center}
\includegraphics[width=6cm,angle=0,clip]{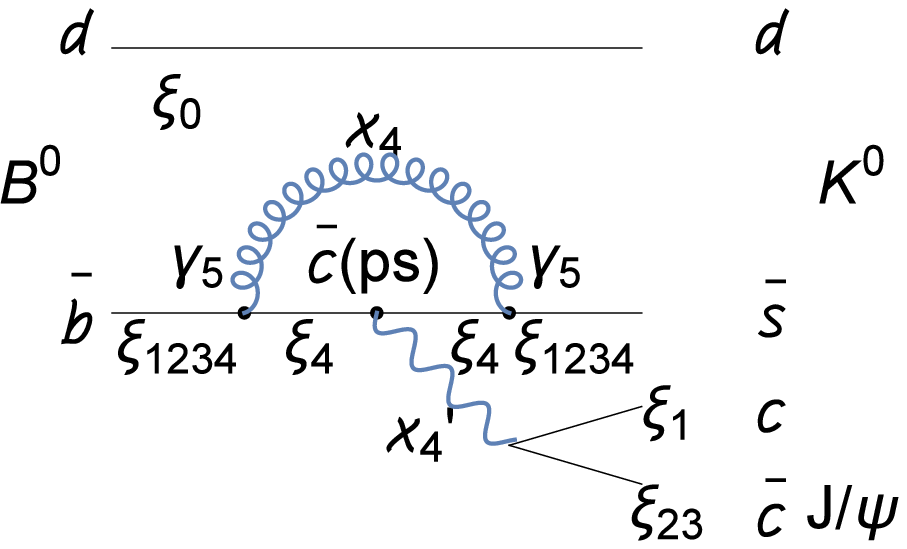} 
\end{center}
\label{bscc551234_4}
\end{minipage}
\hfill
\begin{minipage}[b]{0.47\linewidth}
\begin{center} 
\includegraphics[width=6cm,angle=0,clip]{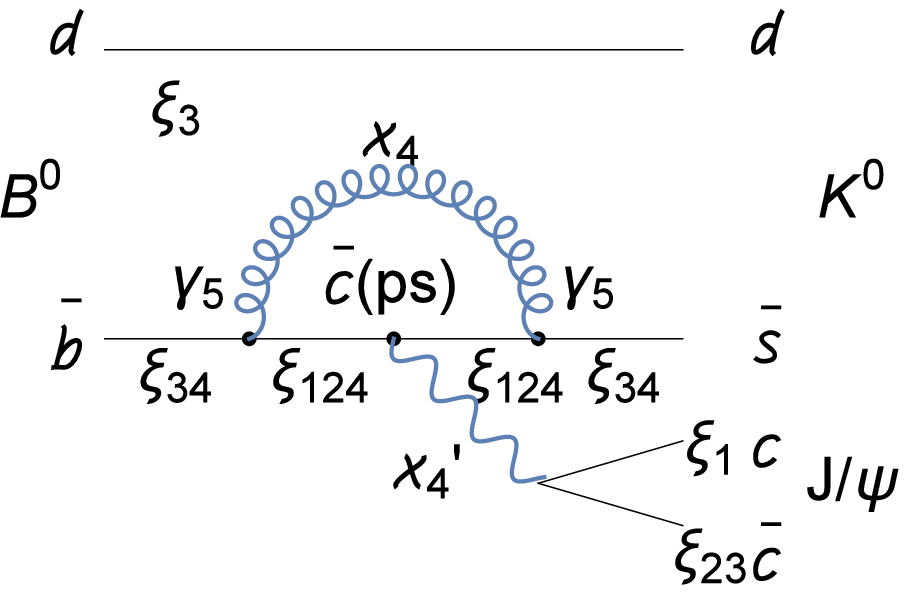} 
\end{center}
\label{bscc5512_3}
\end{minipage}
\begin{minipage}[b]{0.47\linewidth}
\begin{center}
\includegraphics[width=6cm,angle=0,clip]{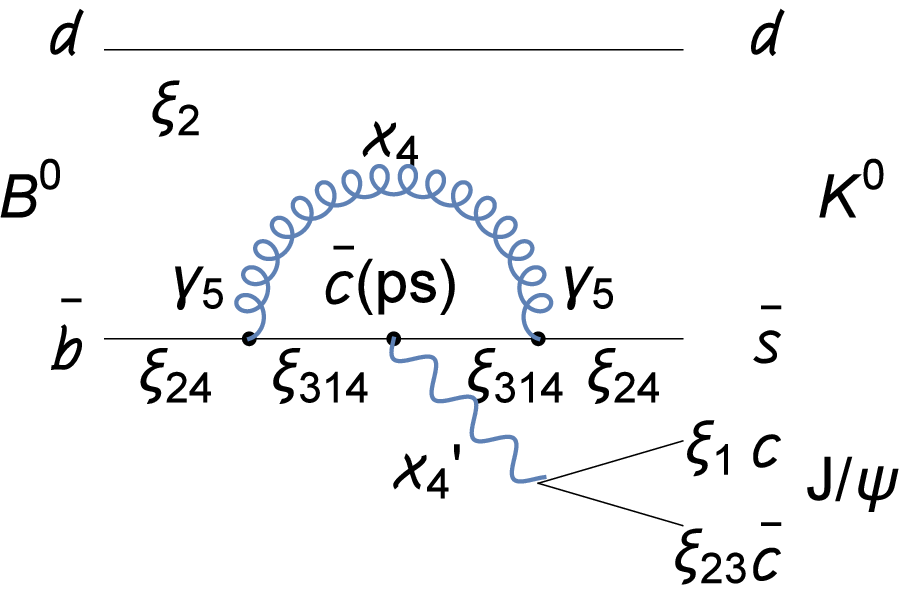} 
\end{center}
\label{bscc5531_2}
\end{minipage}
\hfill
\begin{minipage}[b]{0.47\linewidth}
\begin{center} 
\includegraphics[width=6cm,angle=0,clip]{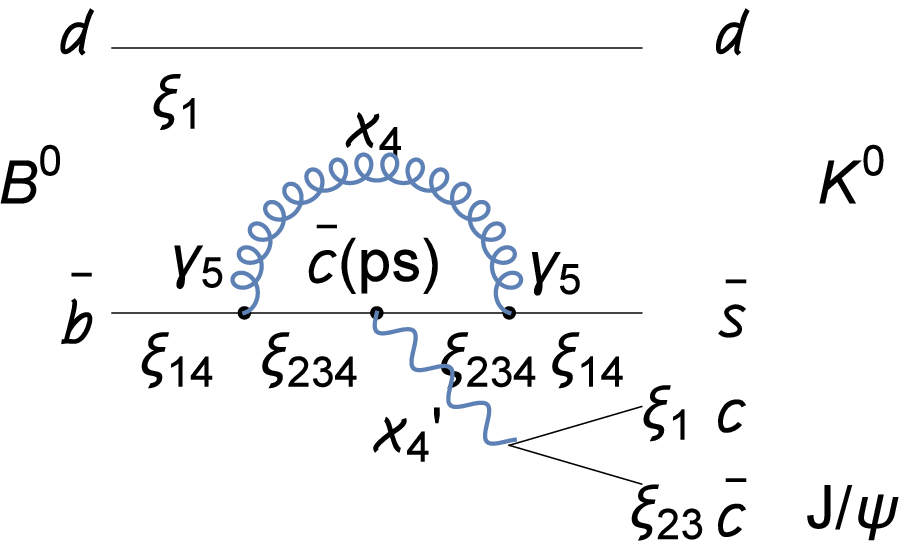}  
\end{center}
\end{minipage}
\caption{Typical diagrams of  $B^0\to K^0\, J/\Psi$ decay, penguin diagram $\gamma_5\gamma_5$ type. In the $J/\Psi$ configurations, ${x_4}'$ decays to $\xi_{1}\xi_{23}$, $\xi_{2}\xi_{31}$ or $\xi_{3}\xi_{12}$.}
\label{bscc55}
\end{figure}

In penguin diagrams of $B^0\to K^0 J/\Psi$, an anti-quark $\bar b$ emits a vector particle $x_4'$ and transforms itself to a $\bar c$ quark, emits a vector particle $x_4'$ that changes to a $J/\Psi$,  and absorbs the vector particle and transforms itself to a $\bar s$ quark.  Couplings of loops of a vector particle and an antiquark can be $1\,1$ type, shown in Fig.\ref{bscc11} and $\gamma_5\gamma_5$ type, shown in Fig.\ref{bscc55}.    
 We can combine the $\gamma_5\gamma_5$ and the $1\,1$ type to make the effective coupling $(1-\gamma_5)(1-\gamma_5)$ type. 

\begin{figure}
\begin{minipage}[b]{0.47\linewidth}
\begin{center}
\includegraphics[width=6cm,angle=0,clip]{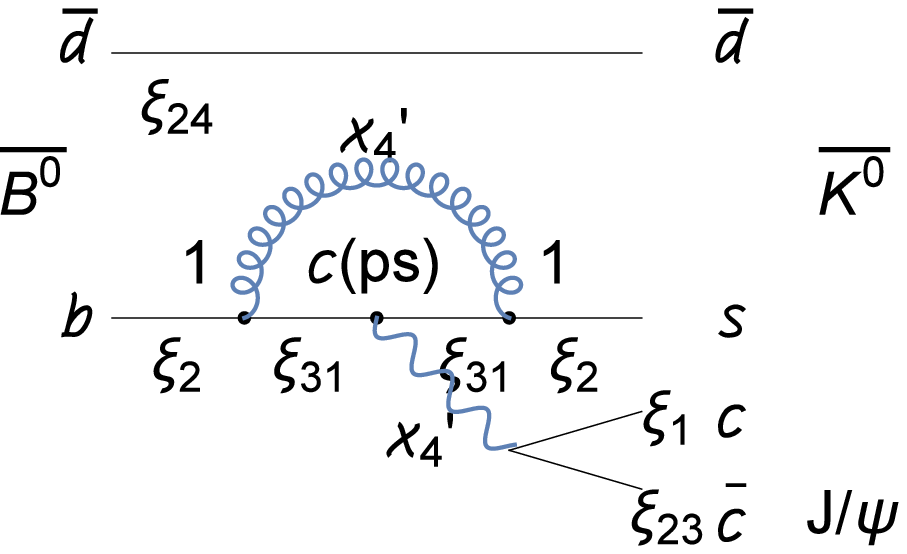} 
\end{center}
\label{bscc22_24}
\end{minipage}
\hfill
\begin{minipage}[b]{0.47\linewidth}
\begin{center}
\includegraphics[width=6cm,angle=0,clip]{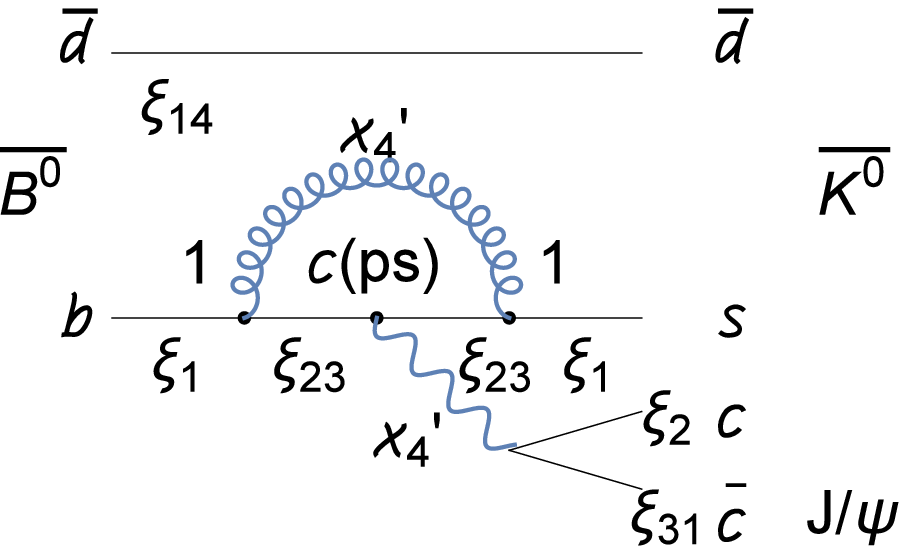}  
\end{center}
\end{minipage}
\caption{Typical diagrams of  $\bar B^0\to \bar K^0\, J/\Psi$ decay. penguin diagram $1\,1$ type. In the $J/\Psi$ final states, a vector particle ${x_4}'$ decays to $\xi_{1}\xi_{23}$, $\xi_{2}\xi_{31}$ or $\xi_{3}\xi_{12}$.}
\label{bscc11_14}
\begin{minipage}[b]{0.47\linewidth}
\begin{center} 
\includegraphics[width=6cm,angle=0,clip]{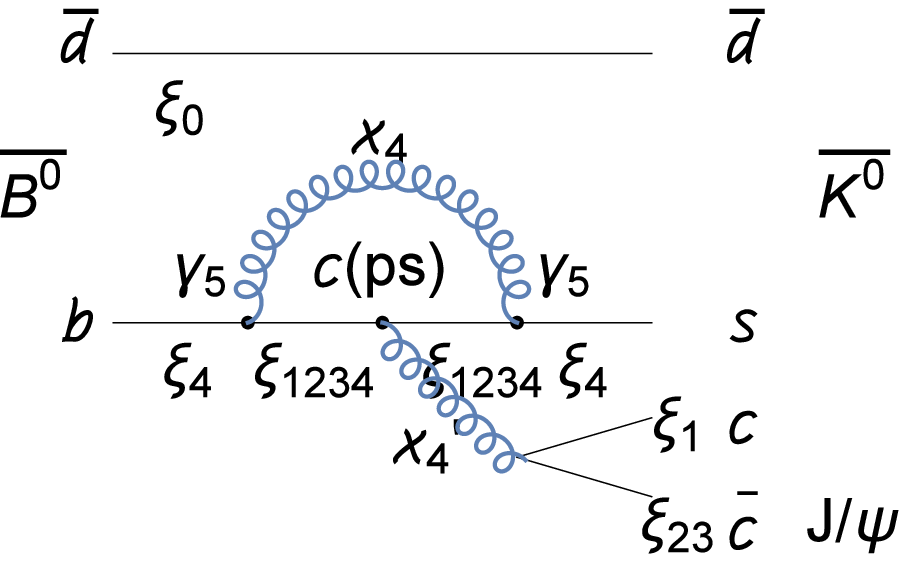} 
\end{center}
\label{bscc11_1234}
\end{minipage}
\hfill
\begin{minipage}[b]{0.47\linewidth}
\begin{center} 
\includegraphics[width=6cm,angle=0,clip]{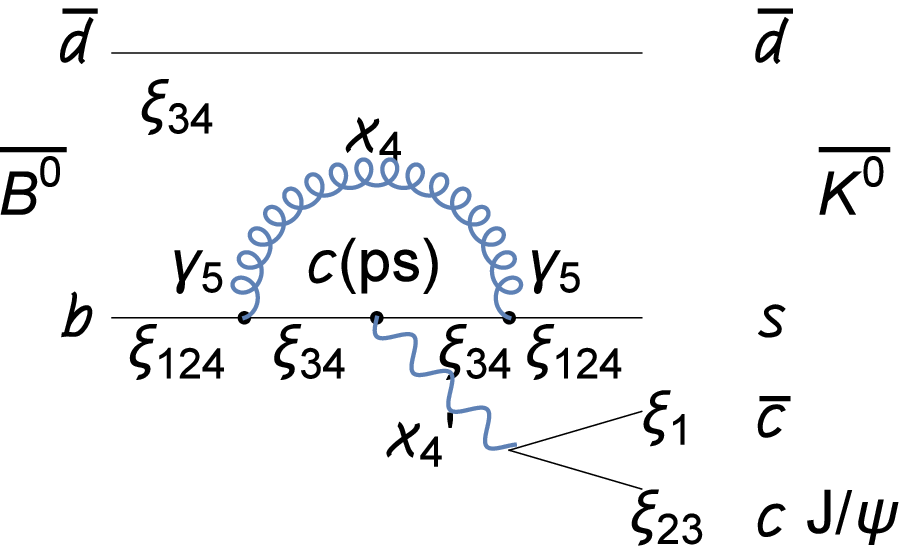} 
\end{center}
\label{bscc11_124}
\end{minipage}
\begin{minipage}[b]{0.47\linewidth}
\begin{center}
\includegraphics[width=6cm,angle=0,clip]{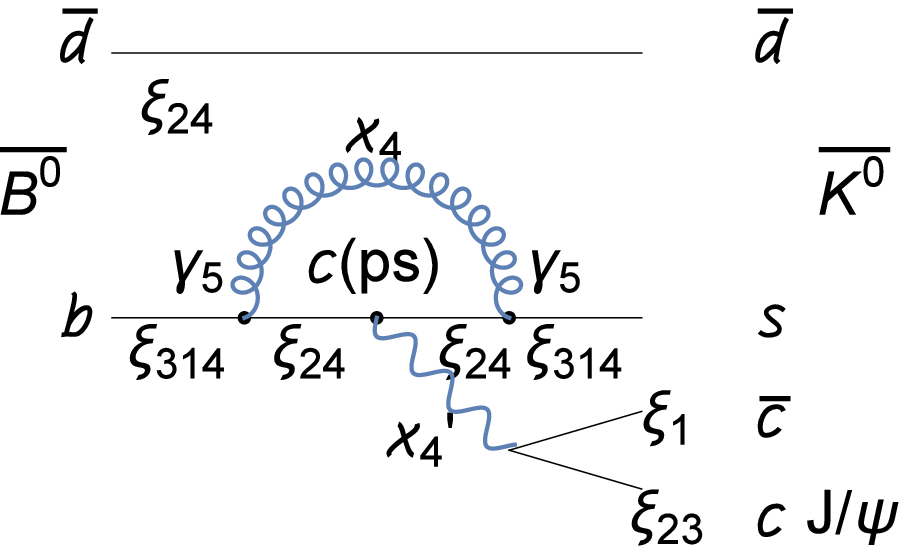} 
\end{center}
\label{bscc11_314}
\end{minipage}
\hfill
\begin{minipage}[b]{0.47\linewidth}
\begin{center}
\includegraphics[width=6cm,angle=0,clip]{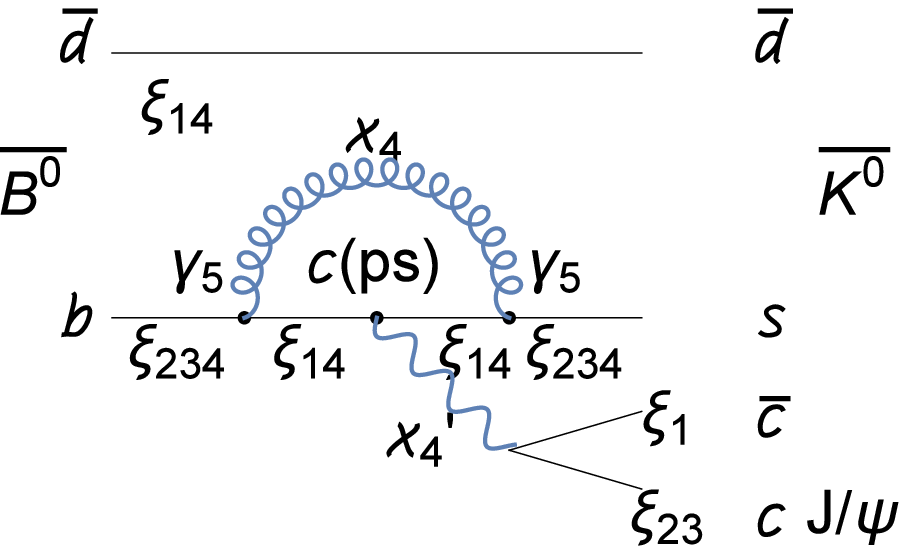}  
\end{center}
\end{minipage}
\caption{Typical diagrams of  $\bar B^0\to \bar K^0\, J/\Psi$ decay, penguin diagram $\gamma_5\gamma_5$ type. In the $J/\Psi$ final state, ${x_4}'$ decays to $\xi_{1}\xi_{23}$, $\xi_{2}\xi_{31}$ or $\xi_{3}\xi_{12}$.}
\label{bscc55_14}
\end{figure}

In penguin diagrams of $\bar B^0\to \bar K^0 J/\Psi$, a quark $b$ emits a vector particle $x_4'$ and transforms itself to a $c$ quark, and emits a vector particle $x_4'$ that changes to a $J/\Psi$,  and absorbs the vector particle and transforms itself to an $s$ quark.  Couplings of a loop of vector particles and an antiquark can be $1\,1$ type, shown in Fig.\ref{bscc11_14} and $\gamma_5\gamma_5$ type, shown in Fig.\ref{bscc55_14}.   The coupling of a $c$ quark and  a vector particle $x_4'$ is pseudoscalar, which we indicate by $(ps)$.

\begin{figure}
\begin{minipage}[b]{0.47\linewidth}
\begin{center} 
\includegraphics[width=6cm,angle=0,clip]{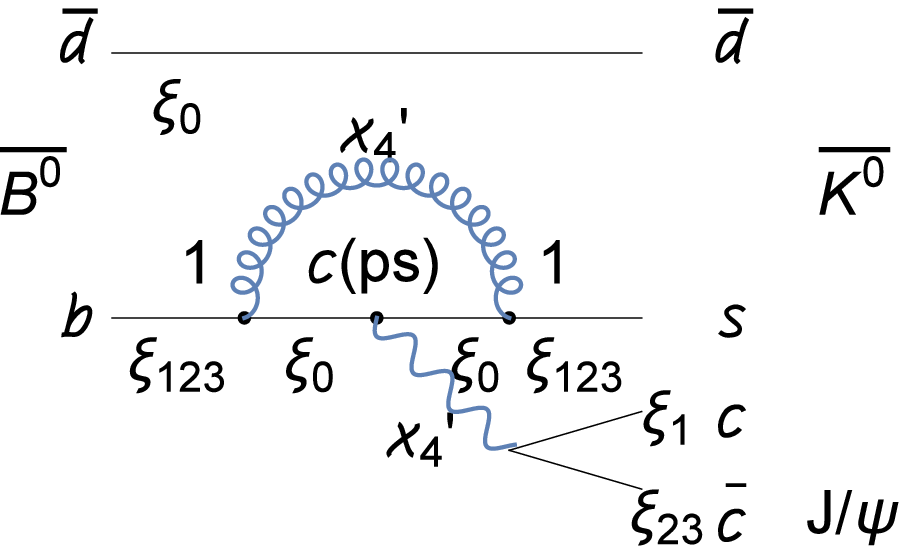} 
\end{center}
\label{bscc123_0}
\end{minipage}
\hfill
\begin{minipage}[b]{0.47\linewidth}
\begin{center} 
\includegraphics[width=6cm,angle=0,clip]{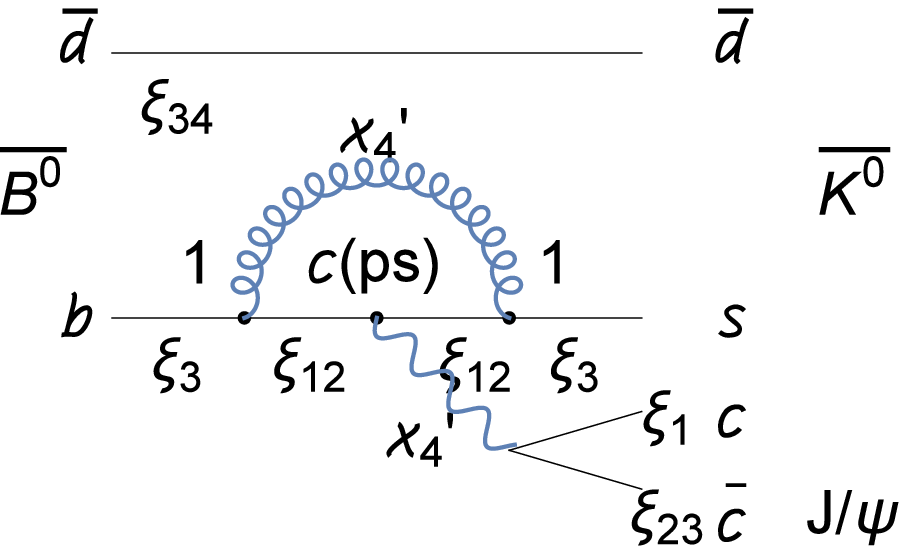} 
\end{center}
\label{bscc33_34}
\end{minipage}
\end{figure}
The penguin diagrams of $\bar B^0\to \bar K^0\, J/\Psi$ of $1\, 1$ type are simple charge conjugation of $B^0\to K^0\,J/\Psi$, but those of $\gamma_5$ type are different. Space components of the $\bar b$ quark in $B^0$ are small components of  Dirac spinors, while those of the $b$ quark in $\bar B^0$ are large components of spinors. It means that amplitudes of $\gamma_5$ contribution in $B^0\to K^0\, J/\Psi$ are suppressed as compared to $\bar B^0\to \bar K^0\, J/\Psi$. 

In the $ B^0\to \bar K^0\,J/\Psi$ decay, there are tree diagrams of type I, in which vector particle of $c\bar c$ is emitted from $\bar b$, or tree diagrams of type II, in which a vector particle that decays to $c\bar s$ is emitter from $\bar s$, and after final state interactions $J/\Psi$ and $K^0$ are created. 

In the analysis of Coulomb interaction between electrons and holes using quaternion basis\cite{FT15}, it was necessary to make couplings of a photon $x_4$ to an electron and a photon ${x_4}'$ to electrons equivalent, for reproducing electron-hole interactions.

 In the present case, couplings of a vector particle $x_4'$ produced from a $\ell\bar\ell$ pair of $J/\Psi$ to a $\bar q$ of $\bar B$, and couplings of a vector particle $x_4$ produced from a $q\bar q$ pair in $J/\Psi$ and a $\bar q$ in $K^0$ are equivalent, and we can create a $\phi\gamma_5{\mathcal C}\psi$ or $\mathcal C \phi\gamma_5\psi$ vertex of $x_4$ from the source of $x_4'$. By the final state interaction, the created $c$ and $\bar c$ interchanged from $\bar b$ in $B^0$ produce $J/\Psi$ meson, and the created $s$ quark and the original $d$ quark of $B^0$ produce a $\bar K^0$ meson. In the evaluation of the matrix elements, we choose the original quark and the original anti-quarks are large components, if it is possible.  The time component of the amplitude of $B^0\to K^0\, J/\Psi$ and $\bar B^0\to \bar{K^0}\,J/\psi$ of $\gamma_5\gamma_5$ type contains the small component $\xi_{1234}$ and they are expected to be much smaller than those of  the $1\,1$ type.

\begin{figure}
\begin{minipage}[htb]{0.47\linewidth}
\begin{center} 
\includegraphics[width=6cm,angle=0,clip]{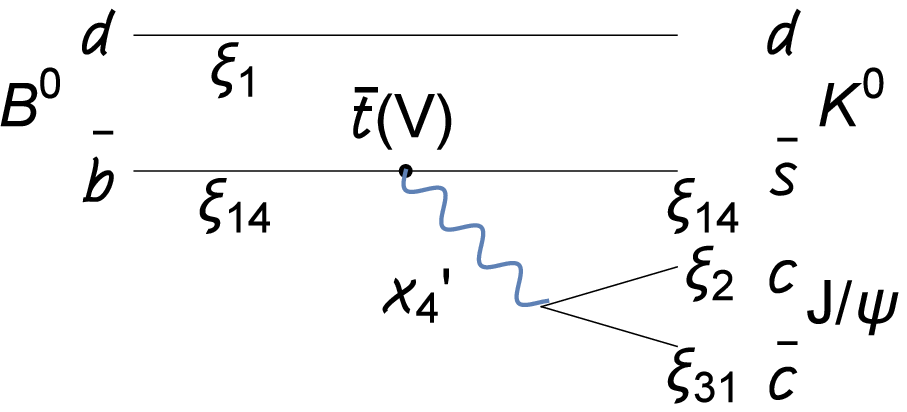} 
\end{center}
\end{minipage}
\hfill
\begin{minipage}[ht]{0.47\linewidth}
\begin{center}
\includegraphics[width=6cm,angle=0,clip]{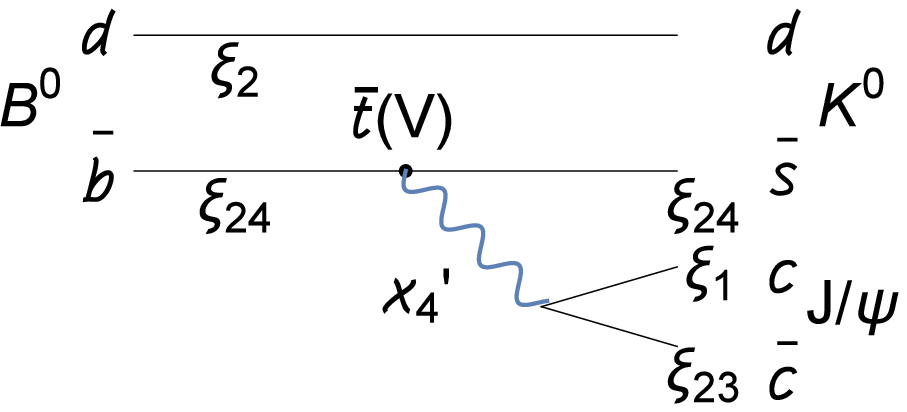}   
\end{center}
\end{minipage}
\begin{minipage}[ht]{0.47\linewidth}
\begin{center} 
\includegraphics[width=6cm,angle=0,clip]{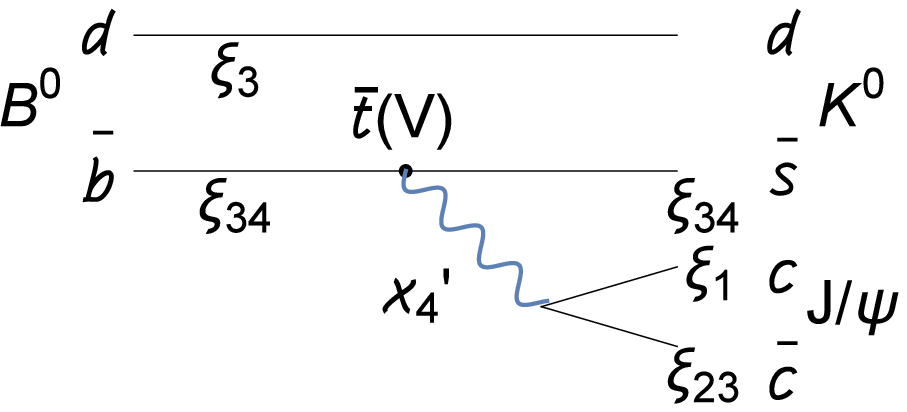} 
\end{center}
\end{minipage}
\hfill
\begin{minipage}[ht]{0.47\linewidth}
\begin{center} 
\includegraphics[width=6cm,angle=0,clip]{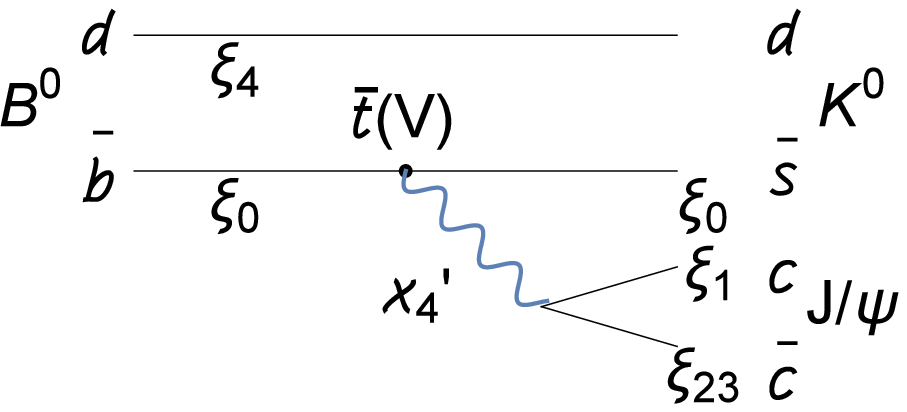}   
\end{center}
\end{minipage}
\caption{Tree diagrams of  $B^0\to K^0\, J/\Psi$ decay (type I). The $J/\Psi$ is produced from the vector particle $x_4'$.  }
\label{bbarT55_11}
\begin{minipage}[htb]{0.47\linewidth}
\begin{center} 
\includegraphics[width=6cm,angle=0,clip]{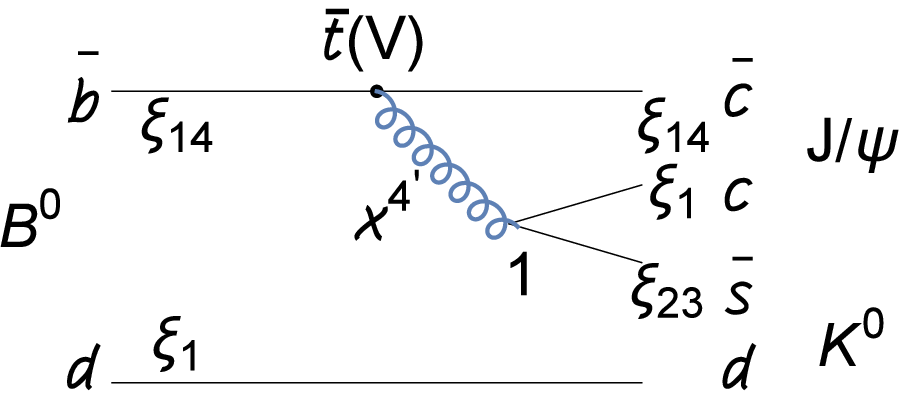}
\end{center}
\end{minipage}
\hfill
\begin{minipage}[ht]{0.47\linewidth}
\begin{center}
\includegraphics[width=6cm,angle=0,clip]{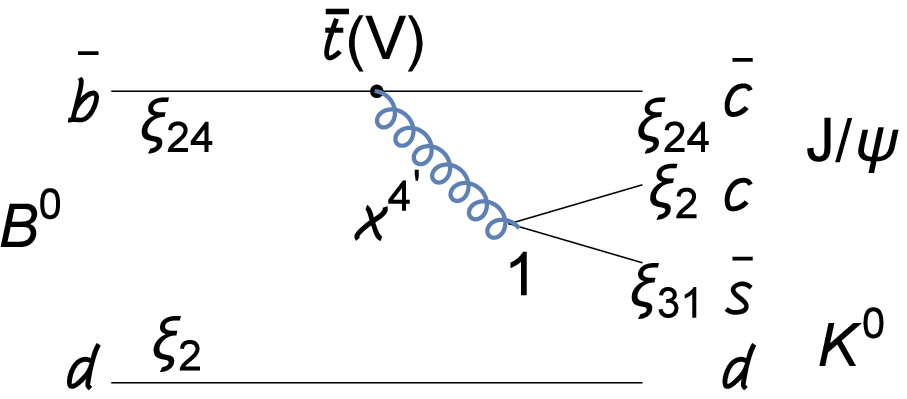}
\end{center}
\end{minipage}
\begin{minipage}[ht]{0.47\linewidth}
\begin{center} 
\includegraphics[width=6cm,angle=0,clip]{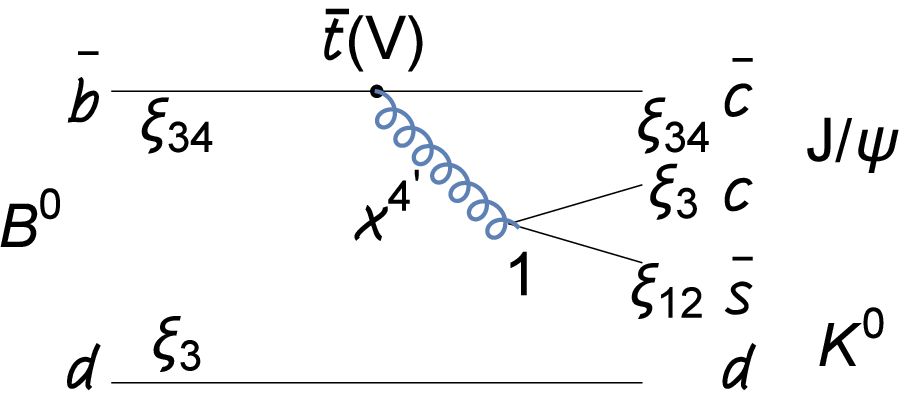}
\end{center}
\end{minipage}
\hfill
\begin{minipage}[ht]{0.47\linewidth}
\begin{center} 
\includegraphics[width=6cm,angle=0,clip]{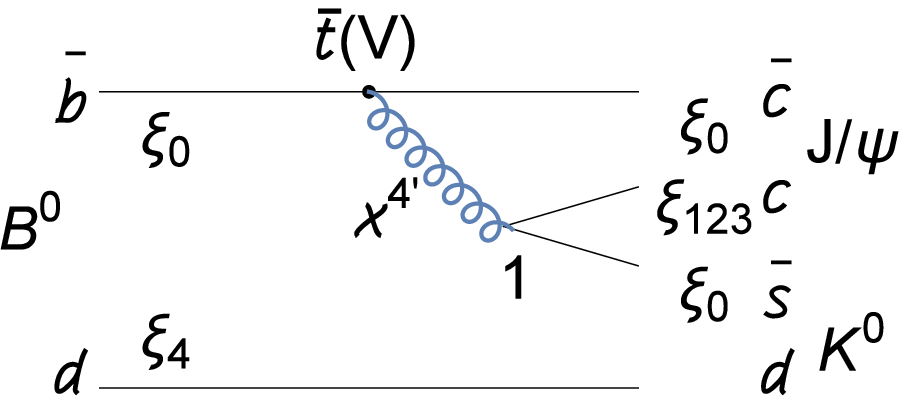}
\end{center}
\end{minipage}
\caption{Tree diagrams of  $B^0\to K^0\, J/\Psi$ decay (type II).  The $J/\Psi$ is not produced from the vector particle $x_4'$.  }
\label{bbardT15_44}
\end{figure}

\begin{figure}
\begin{minipage}[ht]{0.47\linewidth}
\begin{center} 
\includegraphics[width=6cm,angle=0,clip]{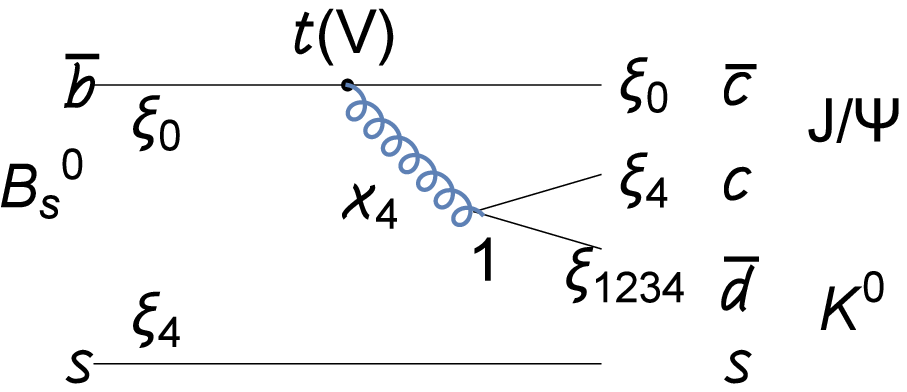} 
\end{center}
\end{minipage}
\hfill
\begin{minipage}[ht]{0.47\linewidth}
\begin{center} 
\includegraphics[width=6cm,angle=0,clip]{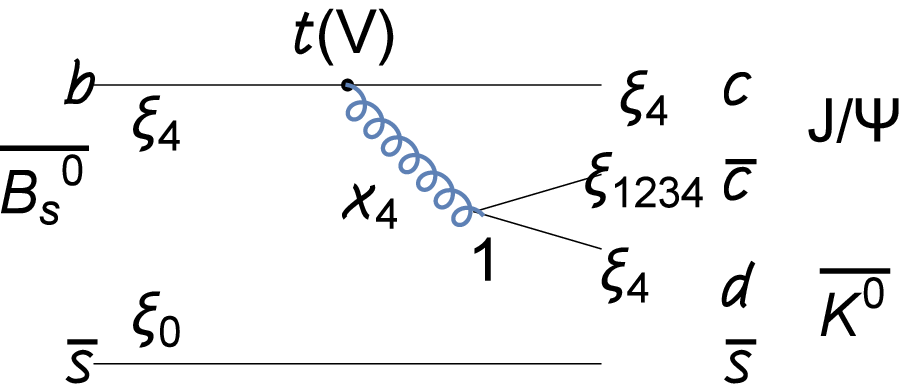} 
\end{center}
\end{minipage}
\caption{Typical tree diagrams of  type II $B^0_s\to K^0\, J/\Psi$ decay (left) and $\bar B^0_s\to \bar K^0\, J/\Psi$ decay (right). }
\label{bdbarT15_44}
\end{figure}

In the case of $1\,1$ type interactions, diagrams of type I is symmetric under charge conjugation, but diagrams of type II $B_s^0\to K^0\, J/\Psi$ and $\bar B_s^0\to \bar K^0\, J/\Psi$ as shown in Fig.\ref{bdbarT15_44} are not symmetric. 
The $1\,1$ interaction of  $B_s^0\to K^0\, J/\Psi$, $J/\Psi$ consists of $\bar c(\xi_0) c(\xi_4)$ and $K^0$ consists of $\bar s(\xi_{1234})d(\xi_4)$, i.e. $K^0$ has the small component $\xi_{1234}$, or the strength of the tree diagram of $c\bar c\,K^0$ becomes weak, same as in penguin diagram of $\gamma_5\gamma_5$ type and the number of difference of events $N_{\bar B_s^0}-N_{B_s^0}$ becomes large.  In the case of space components of  $B_s^0\to K^0\, J/\Psi$, suppression of $J/\Psi$ creation occurs, but it would be difficult to detect from experimental data of $c\bar c\,K^0$. 

The $1\,1$ interaction of $\bar B_s^0\to \bar K^0\, J/\Psi$, $J/\Psi$ consists of $c(\xi_4)\bar c(\xi_{1234})$, and $\bar K^0$ consists of $d(\xi_4)\bar s(\xi_0)$, i.e. $J/\Psi$ has the small component $\xi_{1234}$. Effects of this suppression on the strength of the $c\bar c\, \bar K^0$ may be difficult to detect, but in the case of the space components of $\bar B_s^0\to \bar K^0\, J/\Psi$, suppression of $\bar K^0$ creation occurs, and  becomes weak, same as in penguin diagrams of $\gamma_5$ type, and the number of difference of events $N_{\bar B_s^0}-N_{B_s^0}$ becomes small. The $B^0$ tags and the $\bar B^0$ tags
in  

The above properties are what the BABAR experiment of $c\bar c\, K_L^0$ of $B^0$ tags and $\bar B^0$ tags in the $\Delta t=t_{CP}-t_{flavor}>0$ region show \cite{Babar09}.

The experimental data of $CP$ even final states \cite{Babar09} shows enhancements in raw $CP$ asymmetry in $|\Delta t|\simeq 6$ps region.  We expect this is 
the contribution of type II tree diagram generating $\bar K^0(s(\xi_4)\bar d(\xi_0))$,  and that of $\gamma_5\,\gamma_5$ type penguin diagrams generating $\bar K^0(\bar d(\xi_0) s(\xi_4))$ which is stronger than $K^0(d(\xi_0)\bar s(\xi_{1234}))$.

\begin{figure}
\begin{minipage}[ht]{0.47\linewidth}
\begin{center} 
\includegraphics[width=6cm,angle=0,clip]{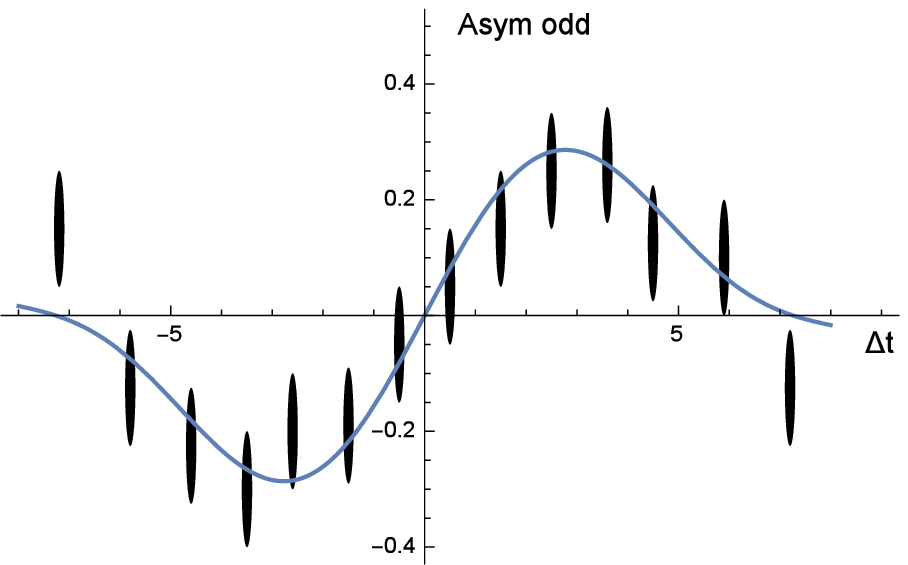}
\end{center}
\end{minipage}
\hfill
\begin{minipage}[ht]{0.47\linewidth}
\begin{center} 
\includegraphics[width=6cm,angle=0,clip]{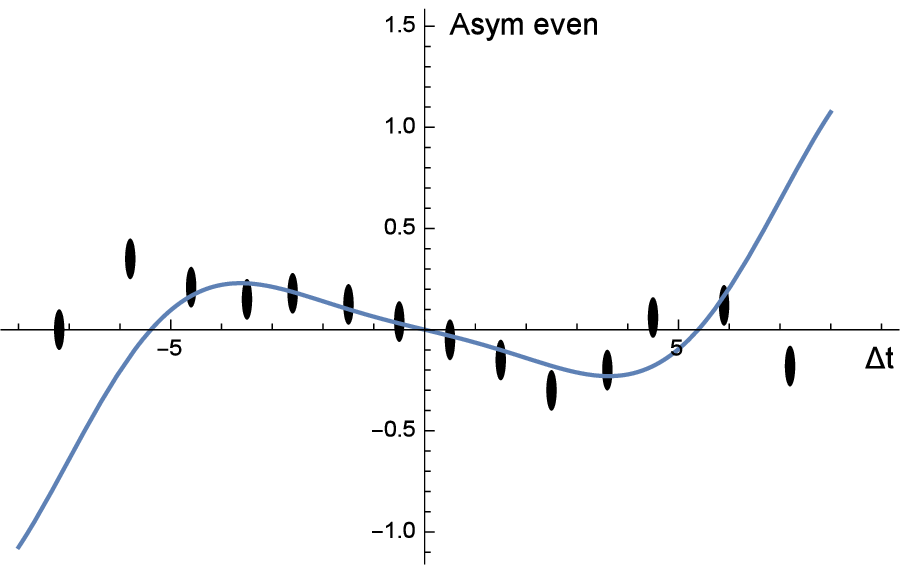}
\end{center}
\end{minipage}
\caption{CP asymmetries in $B^0\to J/\Psi K$ decay of CP-odd final state (left) and CP-even final state (right). }
\label{CPasymmetry}
\end{figure}

 In our $B$ decay to $J/\Psi K$, the CP-odd final state of \cite{Babar09} can be approximated by 
\begin{eqnarray}
&&(e^{(t-1)/16}[1+\cos[\Delta m_{B} (t-2)]))^2-(e^{-(t+1)/16}(1+\cos[\Delta m_{B} (t+2)]))^2\nonumber\\
&&=e^{(t-1)/8} 4\cos^4\frac{\Delta m_B(t-2)}{2}-e^{-(t+1)/8} 4\cos^4\frac{\Delta m_B(t+2)}{2}
\end{eqnarray}
and CP-even final state of \cite{Babar09} can be approximated by 
\begin{eqnarray}
&&\frac{1}{8} (e^{(t-1)/16}(1-\cos[{\Delta m_{B}}(t-2)]))^2-\frac{1}{8} (e^{-(t+1)/16}(1-\cos[{\Delta m_{B}}(t+2)]))^2\nonumber\\
&&=e^{(t-1)/8} \frac{1}{2}\sin^4\frac{\Delta m_B(t-2)}{2}-e^{-(t+1)/8} \frac{1}{2}\sin^4\frac{\Delta m_B(t+2)}{2}
\end{eqnarray}
where $\Delta m_{B}=0.463$ps$^{-1}$ is fixed\cite{GNPS97}, which are shown in Fig.\ref{CPasymmetry}.
Center of an elliptic disk is that of an experimental point, and the error bar of the asymmetrys are fixed to 0.2, and the ratio of error bars in asymmetry axis and in $\Delta t$ axis is fixed  to the goldenratio $(1+\sqrt 5)/2$ which ratio is the value used in Mathematica for presentations of asymmetry curves.

We expect that the experimental asymmetry of $B^0\to  K^0\, J/\Psi$ and $\bar B^0\to \bar K^0\, J/\Psi$ at large $\Delta t$ can be understood from the final state wave functions caused by the $1-\gamma_5$ vertexes.

\section{$B^0$ decay into $D^+D^-$ and $B_s^0$ decay into $D_s^+D_s^-$  }

Cartan's supersymmetry based on Clifford algebra presents large asymmetry between the weak decay of fermions $\psi_L$ and $\mathcal C\phi_L$. 
The origin of the discrepancy between the raw asymmetry of $CP$ even final states $J/\Psi K_L^0$ and the best fit projection in $\Delta t$ of $B^0$ tagged events and $\bar B^0$ tagged events\cite{Babar09} is expected to be due to the $\gamma_5\gamma_5$ type penguin interaction. One can ask whether similar effects can be seen in other decay processes leading to different final states. 

There are analysis of $CP$ symmetry in $B^0\to J/\Psi K_S^0$, $B^0\to D^- D^+$ and
$B_s^0\to D_s^- D_s^+$\cite{Fleischer99}. In the analysis of $B^0\to J/\Psi K_S^0$, penguin diagrams of $\gamma_5$ were not  included and the $CP$ asymmetry was not observed. Results of $B^0\to J/\Psi K_S^0$\cite{Fleischer99,DBF14,Fleischer14}  is less clear than that of $B^0\to J/\Psi K_L^0$.

In the penguin diagram of  $B^0\to J/\Psi \bar K^0$, the vector particle $c\bar c$ emitted from the anti-quark transformed to $J/\Psi$, while in the case of $B^0\to D^+D^-$, the $c$ is absorbed in $D^-$ and $\bar c$ is absorbed in $D^+$. There is not large differences in penguin diagrams and tree diagrams, and clear $CP$ asymmetry was not observed.
An experiment of $B^0\to D^{(*)}\bar{D^{(*)}}$ using partial reconstruction of $B^0\to D^{*-} X\ell^+ \nu_\ell$ and a kaon tag\cite{Babar13} does not show clean $CP$ asymmetry. 
\begin{figure}
\begin{minipage}[b]{0.47\linewidth}
\begin{center} 
\includegraphics[width=6cm,angle=0,clip]{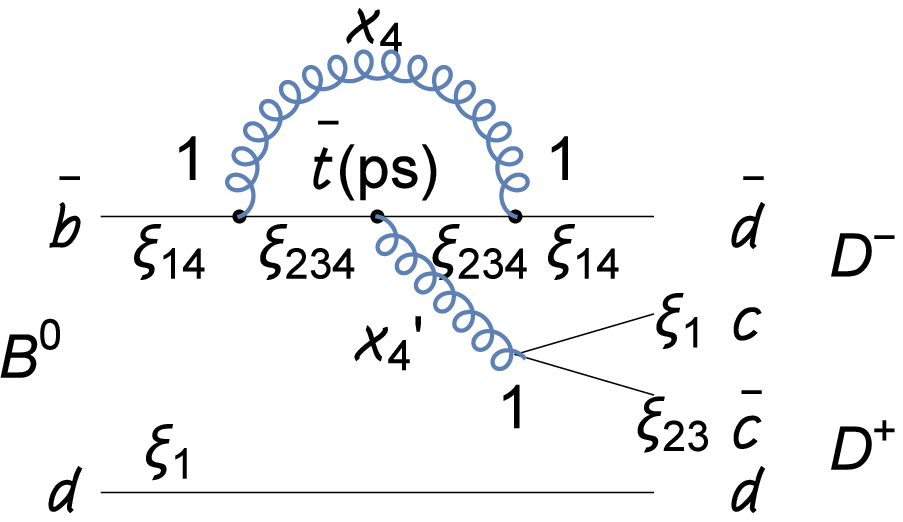}
\end{center}
\end{minipage}
\hfill
\begin{minipage}[b]{0.47\linewidth}
\begin{center} 
\includegraphics[width=6cm,angle=0,clip]{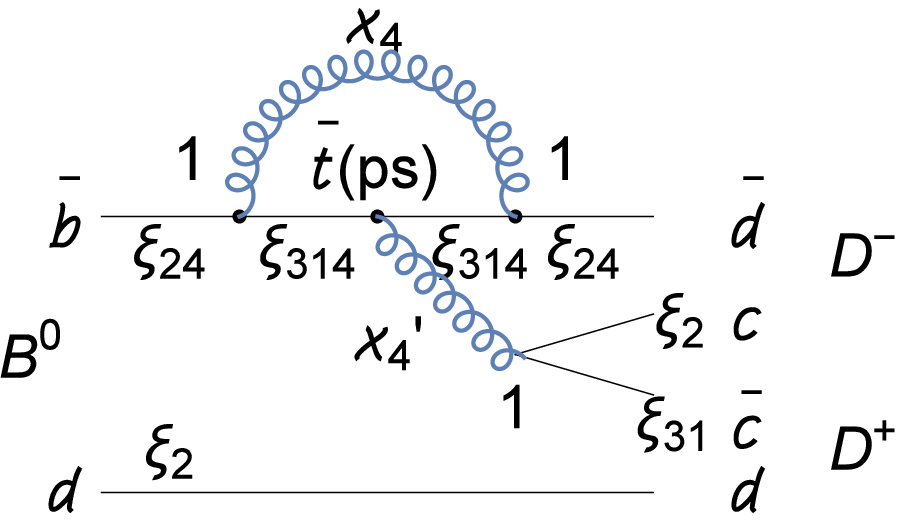} 
\end{center}
\end{minipage}
\begin{minipage}[b]{0.47\linewidth}
\begin{center} 
\includegraphics[width=6cm,angle=0,clip]{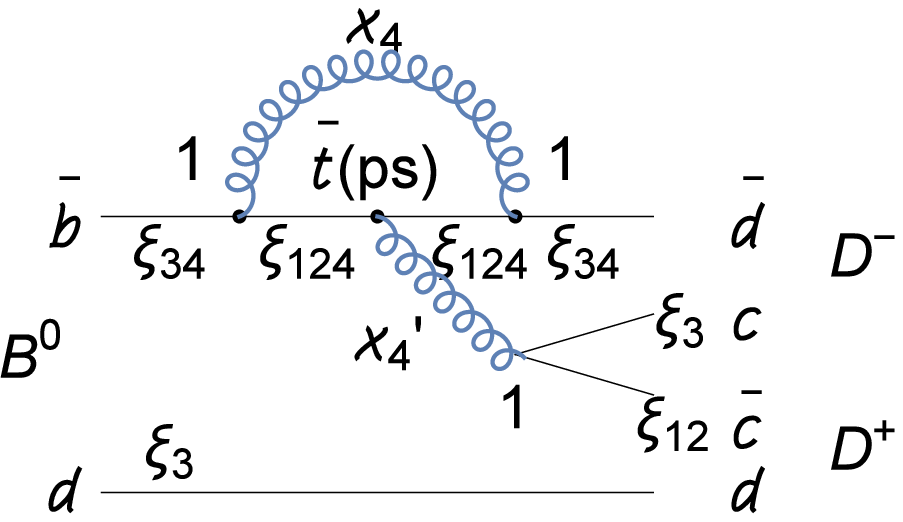} 
\end{center}
\end{minipage}
\hfill
\begin{minipage}[b]{0.47\linewidth}
\begin{center} 
\includegraphics[width=6cm,angle=0,clip]{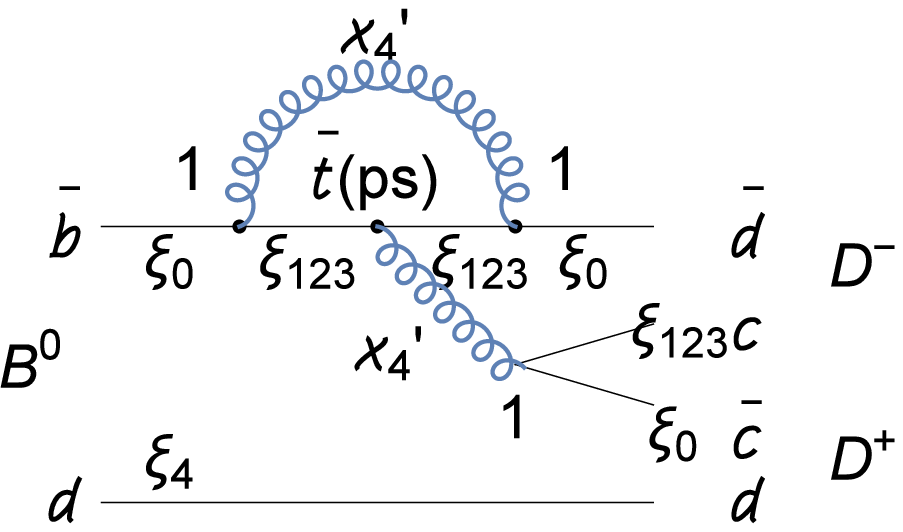} 
\end{center}
\end{minipage}
\caption{Penguin diagrams of  $B^0\to D^- D^+$ decay.}
\label{bdbar55_3p}
\begin{minipage}[b]{0.47\linewidth}
\begin{center} 
\includegraphics[width=6cm,angle=0,clip]{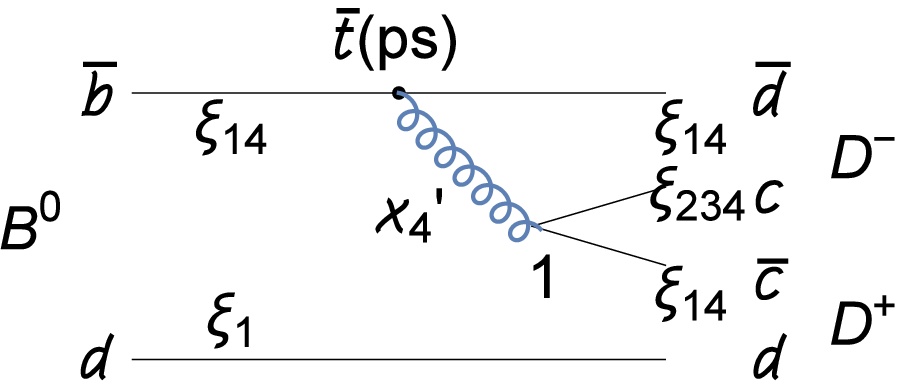} 
\end{center}
\end{minipage}
\hfill
\begin{minipage}[b]{0.47\linewidth}
\begin{center} 
\includegraphics[width=6cm,angle=0,clip]{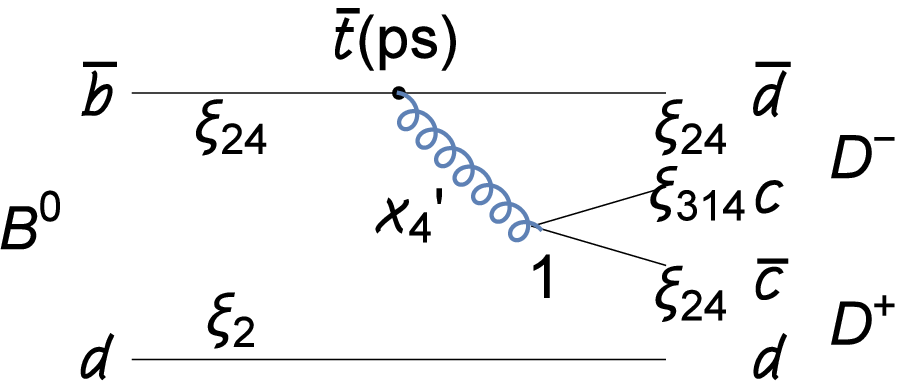} 
\end{center}
\end{minipage}
\begin{minipage}[b]{0.47\linewidth}
\begin{center} 
\includegraphics[width=6cm,angle=0,clip]{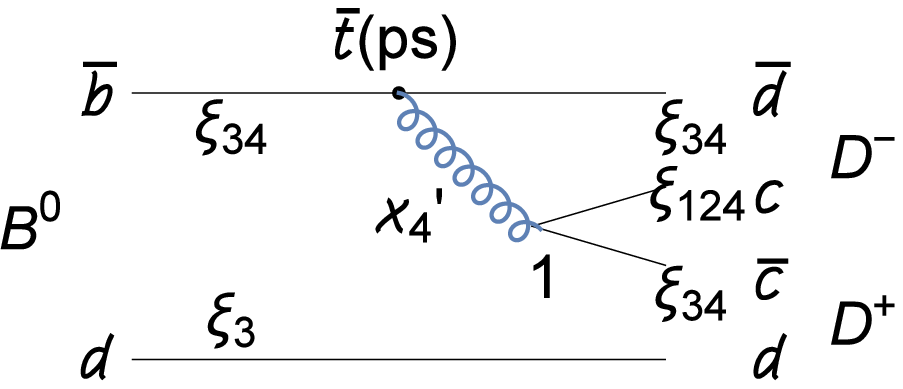}  
\end{center}
\end{minipage}
\hfill
\begin{minipage}[b]{0.47\linewidth}
\begin{center} 
\includegraphics[width=6cm,angle=0,clip]{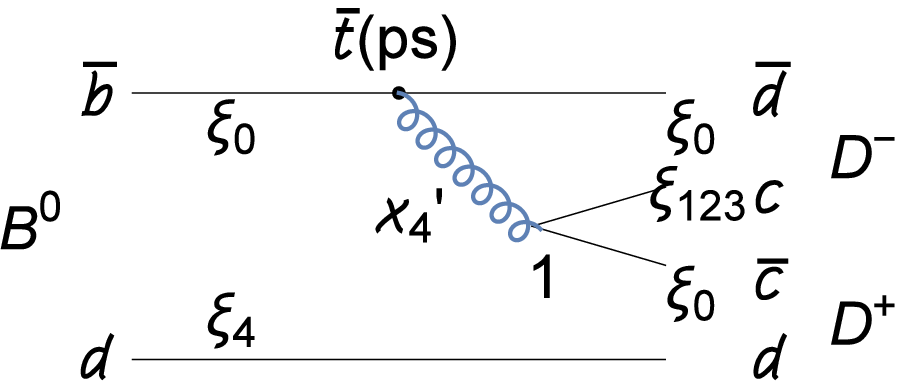} 
\end{center}
\end{minipage}
\caption{Tree diagrams of  $B^0\to D^- D^+$ decay.}
\label{bdbar11_4tt}
\end{figure}

\section{Discussion and conclusion}
We applied Cartan's supersymmetry to weak interaction and analyzed the $B^0\to K^0\,J/\Psi$ and $B^0\to D_d^-D_d^+$ and $B_s^0\to D_s^+D_s^-$ decay using vector particles $x_i$ and $x_i'$ ($i=1,2,3,4$) and assigning couplings to quarks by $1-\gamma_5$ vertices, and inducing effective couplings of $W$ bosons, $Z$ bosons and photons. 

We observed that the $CP$ symmetry of $B^0\to K^0_L J/\Psi$ is violated, while that of   
$B^0\to D_d^- D_d^+$ and $B_s^0\to D_s^- D_s^+$ are not strongly violated. 
In these decay processes, there appear penguin diagrams and tree diagrams, and in the $CP$ asymmetry of $B^0\to K^0_L J/\Psi$ and $\bar B^0\to \bar K^0_L J/\Psi$, space components of penguin diagrams of $\gamma_5\gamma_5$ type and the time component of tree diagrams of $\gamma_5$ type play important roles in changing the strength of $c\bar c\,K^0$ events and $c\bar c\,\bar K^0$ events, where $c\bar c$ is not necessarily the $J/\Psi$ state.

We showed that Cartan's supersymmetry is a useful tool for analyzing violation of $CP$ symmetry of $B^0,\bar B^0$ and $B_s^0, \bar B_s^0$ systems.
Difficulties in choosing appropriate CKM amplitudes $V_{cs}, V_{cb}$ and $V_{cd}$ that explain $B\to\ell\bar\ell$ and $B_s\to\ell\bar\ell$ consistently can be removed by an appropriate assignment of the strength of penguin diagram amplitudes and tree diagram amplitudes \cite{SF15d}.


\end{document}